# Nanoscale characterisation of hydrides and secondary phase particles in Zircaloy-4

Wenjun Lu[1,2], Paraskevas Kontis[1], Siyang Wang[3], Ruth Birch[3], Mark Wenman[3], Baptiste Gault[1,3], T. Ben Britton[3,4*]

1. Max-Planck-Institut für Eisenforschung, Max-Planck-Str. 1, 40237 Düsseldorf, Germany.
2. Department of Mechanical and Energy Engineering, Southern University of Science and Technology, Shenzhen, 518055, China.
3. Department of Materials, Royal School of Mines, Imperial College London, Prince Consort Road, London SW7 2BP, United Kingdom.
4. The Department of Materials Engineering, University of British Columbia, 309-6350 Stores Road, Vancouver, BC Canada V6T 1Z4.

Corresponding author: *ben.britton@ubc.ca

Keywords: hydrogen, zirconium, microscopy

## Abstract

The interaction of hydrogen and metals continues to be industrially relevant and is a critical part of creating and supporting a safety case for nuclear reactor operation. In the present work, we explore hydrogen storage and hydride formation in a zirconium alloy. We characterise the structure and interfaces of fine scale hydrides using scanning transmission electron microscopy (STEM) including energy dispersive spectroscopy (EDS/EDX), electron energy loss spectroscopy (EELS), and high-resolution STEM. Chemical characterisation is supported further with atom probe tomography (APT). Samples were prepared with cryo-focussed ion beam machining (cryo-FIB) and contain hydrides in α-Zr matrix and hydrides associated with one FeCrZr secondary phase particle (SPP). Major findings include characterisation of different interface planes based upon the size of the hydrides and chemical redistribution of solute ahead of the hydride-metal interface. We also find significant (up to 6 at%) hydrogen retained in solution within the zirconium matrix and show a hydride with only 17 at% hydrogen, which is well below that of a ζ-phase stoichiometry suggesting it is an embryonic hydride. These findings help us understand the distribution of hydrogen and the nanoscale morphology of hydrides, which may influence the lifetime of zirconium-based nuclear fuel cladding.

## Introduction

Nuclear power provides 10% of the world's electricity and has avoided about 55 Gt of $CO_2$ emissions over the past fifty years [1]. Of the current operating 442 commercial nuclear reactors, 96% are water cooled. In the majority of these systems, uranium oxide fuel is contained within a zirconium alloy cladding. The cladding acts as the first line of defence to limit fission product release into the primary loop, as well the primary transport medium for neutrons and heat. In these water-based reactors, over time hydrogen may be introduced either directly from the fuel or due to oxidation of the metal, which reduces water and liberates atomic hydrogen [2]. Some of this hydrogen can ingress into the metal and this can impact the performance of the fuel cladding [3–6].

Hydrogen in zirconium has limited solubility, and hydrogen tends to diffuse rapidly from hot to cold regions, as well as towards regions of high tensile stress [7]. When the local region becomes saturated in hydrogen, either due to an increase in hydrogen concentration or due to a reduction in temperature, solid hydrides may form [8–10].

In zirconium-rich polycrystalline alloys, such as Zircaloy-4, the matrix is α-Zr which is a hexagonal (nearly) close packed phase. In this material, hydrides can form inside the α-Zr grains as "intragranular" hydrides. Hydrides can form on grain boundaries as "intergranular" hydrides. Prior work by Wang et al. [11] has highlighted, using electron backscatter diffraction (EBSD), that at the micrometre length scale hydrides form with a $\{0001\}_\alpha ||\{111\}_\delta\; ; <11\bar{2}0>_\alpha\; || <111>_\delta$ orientation relationship (consistent with prior work [12–15]). These macroscopic observations also revealed that the habit plane of the hydride metal interface was consistent often with $\{10\bar{1}7\}_\alpha$ (as reported by Westlake [16]) and that these macroscopic hydrides can often twin along $<111>$, likely due to shear-based mechanisms. Finally, Wang et al. [11] showed that the growth of hydride 'stringers' was likely correlated with a sympathetic growth of hydrides growing from each other.

However, absent from this prior work was an understanding of the chemistry and structure of fine scale intragranular hydrides formed upon rapid cooling [12,17–19] which motivates our present work.

At present there is limited understanding of how the composition of the matrix, with an industrial solubility limit of 1 wppm at room temperature [20] results in the formation of a γ-hydride (10,800 wppm) and limits the creation of physically reasonable models. Lumley et al. [21] showed that for concentrations below 5.9 at% hydrogen, no hydride precipitation is predicted from lattice dynamics simulations which included temperature-dependent vibrational enthalpy and entropy combined with the configurational entropy terms, when not accounting for an nucleation energy barrier. Furthermore, as hydrogen is partially negatively charged when dissolved in the matrix it repels. Experimentally, this present work also builds upon our prior work (Breen et al. [15]) where we used atom probe tomography (APT) and HR-STEM to reveal the structure and chemistry of the (slow cooled) interface of the hydride, which was indexed as δ-phase (FCC) using EBSD. Interestingly, for large hydrides we revealed the presence of a thin (~10 nm thick) HCP ζ-phase at the hydride-matrix interface (the sampling in EBSD is likely to miss this thin phase), as well as rejection of Sn from the hydride enriching the α-Zr/ζ-hydride interface and stacking faults associated with hydride precipitation [22] Recent work by Jia et al. [23] used X-ray diffraction (XRD), scanning electron microscopy (SEM), EBSD and TEM to characterise 'hydride bumps' formed during electrochemical charging of commercially pure zirconium and showed that large >20µm hydride bumps were circular discs or needles once there was sufficient hydrogen introduced into the material, and these discs are lenticular with alignment along the basal plane and with a transformation sequence of Zr → γ-ZrH → δ-ZrH$_{1.66}$. The surface bumps observed were created either from clusters of lenticular subsurface hydrides of mixed twin relationships with an orientation relationship of $(0001)_\alpha||(11\bar{1})_\delta$ or triples of three hydride variations according to the orientation relationship of $(0001)_\alpha||(001)_{\gamma/\delta}$.

The structure, phase and interface of the hydrides is important firstly as they may control the mechanical properties, but also the evolution of the hydrides during thermal cycling can result in a change in performance during life of the clad. Here the role of the hydride phase, their size and distribution, hydride/metal interface, and stored strain energy can affect the dissolution and precipitation during thermal cycling.

For the chemical analysis of the hydrides, this is challenging in APT. Hydrogen is often found within the chamber and the amount of hydrogen ions determined is dependent on the analysis conditions. Furthermore, during focused ion beam sample preparation of the group 4 elements, hydrogen can be introduced if the sample is prepared at room temperature (and this is likely often mis-interpreted as a 'FCC-like' phase [24]). To address these challenges, Mouton et al. [25] and Breen et al. [15] studied this through the use of deuterium charged samples and final polishing at cryogenic temperatures, and

Mouton et al. developed a deconvolution algorithm that can be used to separate the different charge to mass peaks originating from differently ionized Zr, ZrH, and $ZrH_2$ species.

The secondary phase particles (SPPs) in Zircaloy-4 have been studied previously by Annand et al. [26] and they revealed the presence of $Zr(Fe,Cr)_2$ and $Zr_2Fe$ particles (and studied their impact on corrosion under pressurised water conditions), similar to other work in this field. For the present work, we focus on the $Zr(Fe,Cr)_2$ precipitate.

Meng and Northwood [27] found these Laves phase precipitates with a C14 hexagonal structure with a narrow composition range of Cr:Fe of (0.55-0.57):(0.45-0.43), in Zircaloy-2, with a size of around 300 nm containing a stacking fault structure which appear to have a approximately round structure in their low magnification bright field TEM micrographs. The chemical composition was found to be $Zr(Fe_{0.58},Cr_{0.42})_2$ from electrochemical extraction of precipitates from Zircaloy-4 by Toffolon-Masclet et al. [28] and they used X-ray diffraction to measure the lattice parameters of the metastable $Fe(Cr,Fe)_2$ phase C15 (Fd3m space group) precipitates as a=0.7203±0.0003 nm. In a different study by Meng and Northwood [29] used high order Laue zone analysis from selected area diffraction patterns to identify $Zr(Fe,Cr)_2$ precipitates as C15 type cubic Laves phases, and argue due to extinction rules associated with the zero and non-order Laue reflections that their precipitates cannot be C14 hexagonal. The role of SPPs on the oxidation behaviour of the Zircaloy has been studied in the work by Annand et al.[26] where a bulk SPP (in the metal and 2.8 um from the oxide/metal interface) have a faceted outline with a Fe/Cr ratio of 0.83 and segregation of Fe towards the centre of the precipitate. Density functional theory simulations by Burr et al. [30] focussed on binary $ZrCr_2$ and $ZrFe_2$ cubic and hexagonal structures and revealed that they tend to have unfavourable interstitial sites of hydrogen defects, as compared to the $\alpha$-Zr matrix. Recently Jones et al. [31] have shown that hydrogen tends to segregate to the SPP interface using NanoSIMS characterisation and further density functional calculations of ternary Zr-Fe-Cr compositions showed that hydrogen solution enthalpy was still positive relative to the $\alpha$-Zr matrix for all SPP tetrahedral sites.

There has been limited reported literature of hydrides near SPPs. Blat et al. [32] observe large hydrides branching from $Zr(Fe,Cr)_2$ precipitates but these tend to be large with a thickness equal to the diameter of the SPP, in high hydrogen concentration charged (400 ppm) Zircaloy-4. Zanellato et al. [33] observed that C15 precipitates stayed during *in-situ* heating of Zircaloy-4, with thermal cycling to dissolve and precipitate hydrides, and for high hydrogen concentration (475 and 600 ppm) samples the hydrides dissolved fully at 550°C. In Zanellato's work, the hydrides are identified as δ from the presence of the {311}, {022}, {002} and {111} X-ray peaks when the sample is subject to 5°C/min heating and cooling. Notably, for their kinetic analysis they assume that the concentration of hydrogen in the matrix at 50°C is 0 ppm. At a faster cooling rate (above 20°C/min), Zanellato et al. reveal a hysteresis of ~80°C between the terminal solubility temperature (TSS) for dissolution during heating and precipitation during cooling. Notably, they also discuss the potential that the hydrides are under strain at room temperature, but due to uncertainty in the experimental set-up they comment that they cannot draw conclusions from their analysis.

In the present work, we explore the structure and chemistry of nm-sized intragranular hydrides, specifically revealing their structure, chemistry, and morphology using a combination of microscopy techniques (TEM, STEM, EELS and APT). These hydrides are likely newly formed, which enables us to explore the concentration of hydrogen near a newly formed hydride. As the literature tends to focus on role of SPPs on hydrogen oxidation and hydrogen pick-up, we also explore hydride precipitation near a SPP.

## Materials & Methods

A sample of as-received rolled and recrystallised Zircaloy-4 plate was cut. The material has a nominal chemical composition of Zr-1.5%Sn-0.2%Sn-0.1%Cr (wt%) [34]. This was annealed at 800°C for two weeks to grow 'blocky alpha', as per the recipe developed by Tong and Britton [35]. Hydrogen charging was performed as per Ref.[36]. This involved electrochemical hydrogen charging in a solution of 1.5 wt% sulphuric acid using a current density of 2 kA/m$^2$ at 65°C for 24 hr. After charging, the sample was annealed at 400°C for 5 h to homogenise the hydrogen distribution and subsequently quenched in water to room temperature to promote the formation of fine scale hydrides. The sample was mechanically polished with colloidal silica and electropolished with 10 vol% perchloric acid in methanol at -40°C for 90s, applying a voltage of 30V.

The sample was imaged with a Zeiss Merlin SEM, using backscatter electron imaging at high magnification and optimising the electron imaging conditions to reveal a fine dispersion of intragranular hydrides (Figure 1), which decorate the grain interior far from the large scale intergranular hydrides. This approach is close to the electron-channelling contrast imaging (ECCI), albeit not performed with orientation-control [37].

In order to avoid introducing hydrogen and result in the formation of artificial hydrides, the APT and TEM specimens were prepared at cryogenic conditions at -193 °C. In particular, site specific lift-outs from inside an α-Zr grain that contains the intragranular hydrides were performed following the procedures described in Ref [38]. A dual beam SEM / FIB FEI Helios 600 equipped with a GATAN C1001 stage was used for the sharpening of APT specimens at -193°C, by circulation of nitrogen gas cooled by liquid nitrogen using a setup described in [39] . The importance of cryo polishing step for these samples is highlighted in Ref [40]. The sharpening of APT specimens was followed by a warm up of the stage to room temperature and subsequent transfer for atom probe analysis through air. The transfer time between the FIB and the atom probe stage maintained at 60K was minimised, and lasted less than approx. 4h. Specimens were analysed on a Cameca LEAP 5000 XS instrument operating laser pulsing mode with a repetition rate of 250kHz, pulse energy 60pJ and at a base temperature of 60K. Data reconstruction and processing was performed using the Cameca IVAS 3.8.4 software tool.

In the case of the TEM sample preparation, the same FIB equipped with the GATAN C1001 stage was used for thinning the sample. Bright-field TEM imaging as well as selected area electron diffraction (SAED) were carried in an image probe aberration-corrected FEI Titan Themis 80-300 (Thermo Fisher Scientific) microscope operated at 300 kV with semi-convergence angle of 17 mrad. For high-angle annular dark field (HAADF) and low angle annular dark field (LAADF) imaging, we used inner and outer semi-collection angles ranging from 73 to 200 mrad and 14 to 63 mrad, respectively. EELS spectra were recorded using a Quantum ERS (Gatan) energy filter in the image-coupled mode with an entrance aperture collecting electrons scattered up to 35 mrad. Multivariate analysis was performed on the hyperspectral datasets to separate spectral contribution from the two phases (hydride and matrix).

## Results

### Hydride Distributions

ECCI imaging in Figure 1 reveals the presence of many fine scale hydrides decorated inside the α-Zr matrix.

The intragranular metastable hydrides are small and thin. In these 2D images, they typically are 50-250 nm in length and up to <30 nm wide (i.e. a few pixels). The majority of these hydrides are linear and parallel, but even at this low magnification (as compared to the transmission electron micrographs shown later) some hydrides can be seen to be made up of several connected linear segments. The contrast between these hydrides and each matrix grain varies, likely due to the change in incoming channelling vector with respect to the orientation of both grains (and the position of the backscatter detector).

The intergranular hydrides are longer and wider and decorate most but not all grain boundaries. Near the grain boundary hydrides there is an absence of intragranular hydrides, likely because the hydrogen from the matrix was absorbed by the growing hydride. This hydride-denuded zone extends to approximately the width of the intergranular hydride zone. As seen in Figure 1a, within an apparently continuous intergranular hydride, there are interruptions, each of which are nearly parallel. The intragranular hydride decorates one side of the grain boundary in preference to another (Figure 1b) and each tip of the hydride is sharp. One tip of the hydride can extend away from the grain boundary itself.

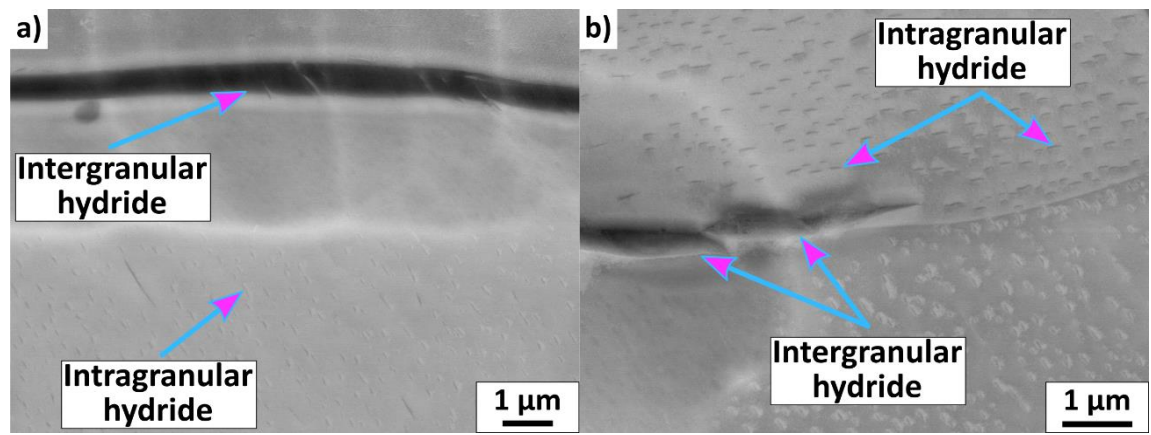

Figure 1: Electron channelling contrast imaging of hydrides in Zircaloy-4

Using cryo-FIB preparation, a TEM foil was extracted from the matrix region, containing intragranular hydrides and a SPP. The region has uniform thickness and has been capped with platinum to protect the region of interest during sample preparation. Within this foil, the size, distribution and extent of the hydrides was studied in more details. Bright-field TEM, HAADF- and LAADF-STEM images (Figure 2) with the SAED reveal the orientations, orientation relationships, and the interface planes of these hydrides and the zirconium matrix.

### Hydride Analysis via Imaging and Diffraction

The hydrides consistently show the $<110>_{FCC}||<11\bar{2}0>_{HCP}$ and the $\{111\}_{FCC}||\{0001\}_{HCP}$ orientation relationship. Within the population of intragranular hydrides imaged in Figure 2, we observed many near vertical hydrides (Type 1) which have an $\{111\}_{FCC}||\{0001\}_{HCP}$ long axis and interface (see the hydrides marked with red lines in Figure 2d). There are longer hydrides that consist of regions which have a Type 1 interface $\{0001\}_{HCP}$ and long axis for some of their length (the blue line in Figure 2d). There are regions where the hydrides deviate from this morphology with an interface that consistent with an interface plane trace of $\{110\bar{1}\}_{HCP}$. Furthermore, there are also regions where two small Type 1 hydrides appear joined by an intermediate region of Type 2 hydride which is aligned with $\{1\bar{1}0\bar{7}\}_{HCP}$ (the green line in Figure 2d).

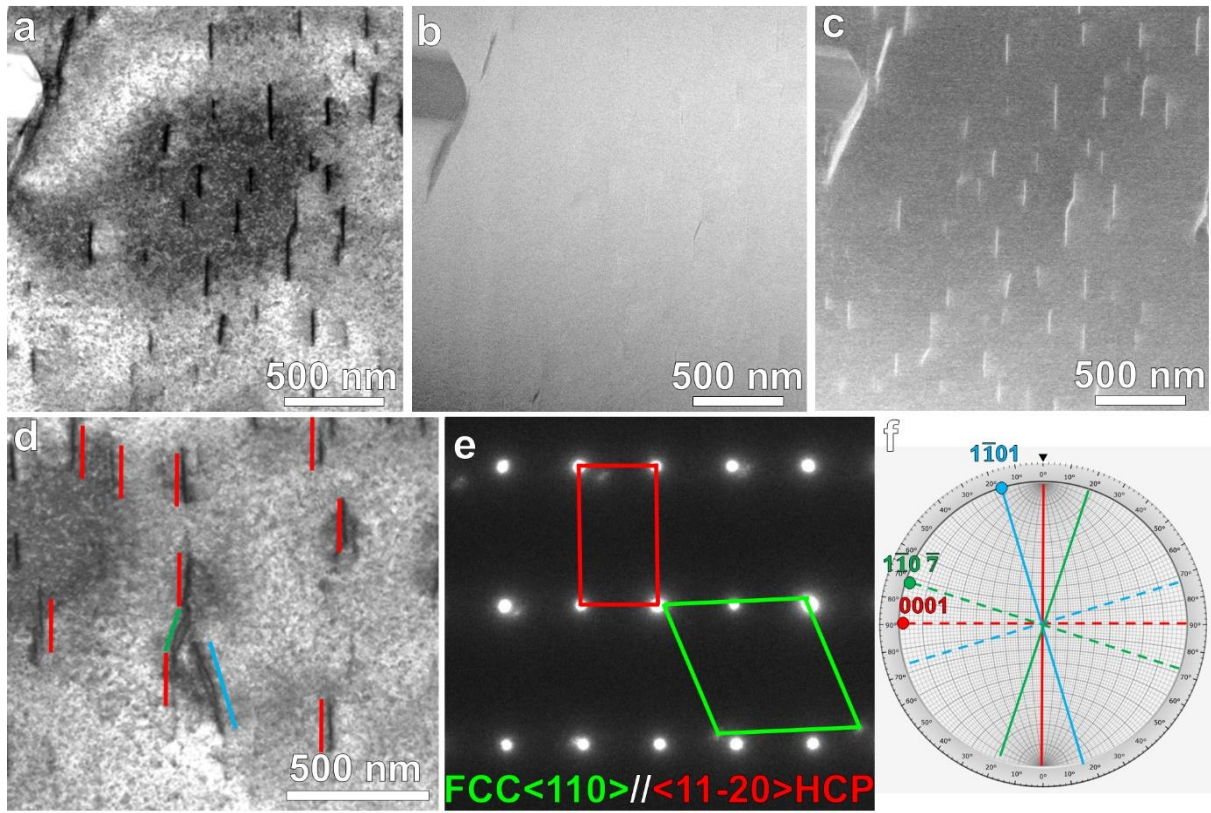

*Figure 2: Intragraular hydrides characterized by (a) bright-field TEM, (b) HAADF-STEM and (c) LAADF-STEM. (d) The macroscopic habit planes are highlighted by solid red, green and blue lines in the bright-field TEM image. (e) SAED pattern of this region. (f) Analysis of these planes with FCC structure by standard stereographic projection identifies them as part of the Type 1 interface {0001} and Type 2 interface {1̄10̄7̄} planes.*

Higher magnification HAADF-STEM imaging of the Type 1 hydride is shown in Figure 3. These STEM images reveal that the hydride is less than 10 nm thick and not continuous, but where the hydride is broken up, the two regions of the hydride are located directly in line with each other. Again, the orientation relationship and the major interfacial plane are consistent with previous observations, but we note that none of these interfaces are sharp.

### Type 1 Hydride Identification

STEM-EELS spectrum imaging of the low-loss region was performed to determine the chemical signature of both Type 1 hydride and matrix, as shown in Figure 3c and d. The overlaid phase maps indicate that the hydrides can be clearly distinguished from the matrix based on the position of their plasmon peaks. The plasmon peak energy is shifted from 16.9 eV in the HCP matrix to 18.4 eV in Type 1 hydride. Based on the literature data [41], the Type 1 hydride is determined to be γ hydride.

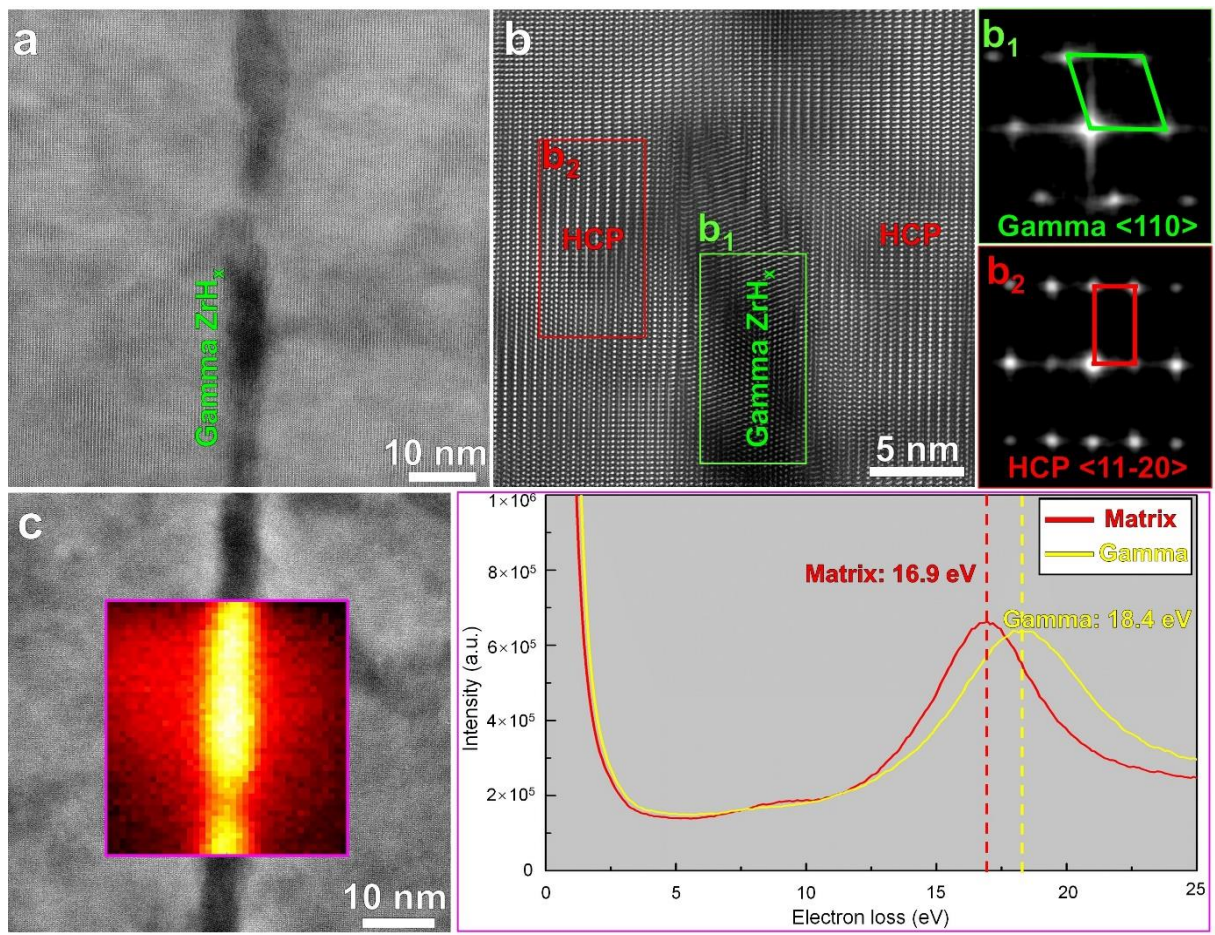

*Figure 3: STEM/EELS analysis of the cyro-FIB sample after H-charging. The needle-shaped hydride along {0001} habit plane (Type 1) is characterized at (a) low and (b) high magnification by HAADF-STEM. ($b_1$ and $b_2$) present the corresponding FFTs marked in (b). (c) Area of EELS spectrum imaging overlaid phase map (yellow: hydride, red: matrix). and ELS spectra in the low energy loss region from the respective areas show their plasmon peak features at different energy loss. From this, the Type 1 hydride is determined to be γ hydride.*

### Type 2 Hydride Identification, linking Type 1 Hydrides

Higher magnification HAADF-STEM imaging and analysis of a hydride containing both the Type 1 and Type 2 interface is shown in Figure 4. The Type 1 domain is thinner and terminates at a sharp point (towards the top and bottom of the hydride in the HAADF and LAADF images in Figure 2). The Type 2 region connects two Type 1 domains. The lower Type 1 region extends upwards in and beyond the thicker Type 2 region. This morphology indicates that this Type 1 region may have existed prior to the growth of the Type 2 hydride connecting region. Analysis of the atomic column STEM images using 2D Fast Fourier Transform (FFT) based pseudo diffraction highlights that the interfacial plane analysis shown in Figure 2 is consistent, for all regions of the hydrides. This indicates that there is a single orientation of the hydride for this entire domain and this orientation is related via the $<110>_{FCC}||<11\bar{2}0>_{HCP}$ and the $\{111\}_{FCC}||\{0001\}_{HCP}$ orientation relationship, as noted previously. The interface of the Type 2 connecting region is less well defined than the Type 1 region within the hydride. For these images in Figure 4 there is limited evidence of any dislocations associated with these fine scale hydrides.

The STEM-EELS results (Figure 4g) reveal that the Type 2 hydride has a plasmon peak energy 19 eV, consistent with the plasmon peak value of δ-hydride [41]. which is higher energy than that of Type 1 γ-hydride (plasmon peak at 18.4 eV). Hence, the plasmon peak analysis clearly indicates that the Type 1 and Type 2 hydrides are γ and δ-hydrides, respectively.

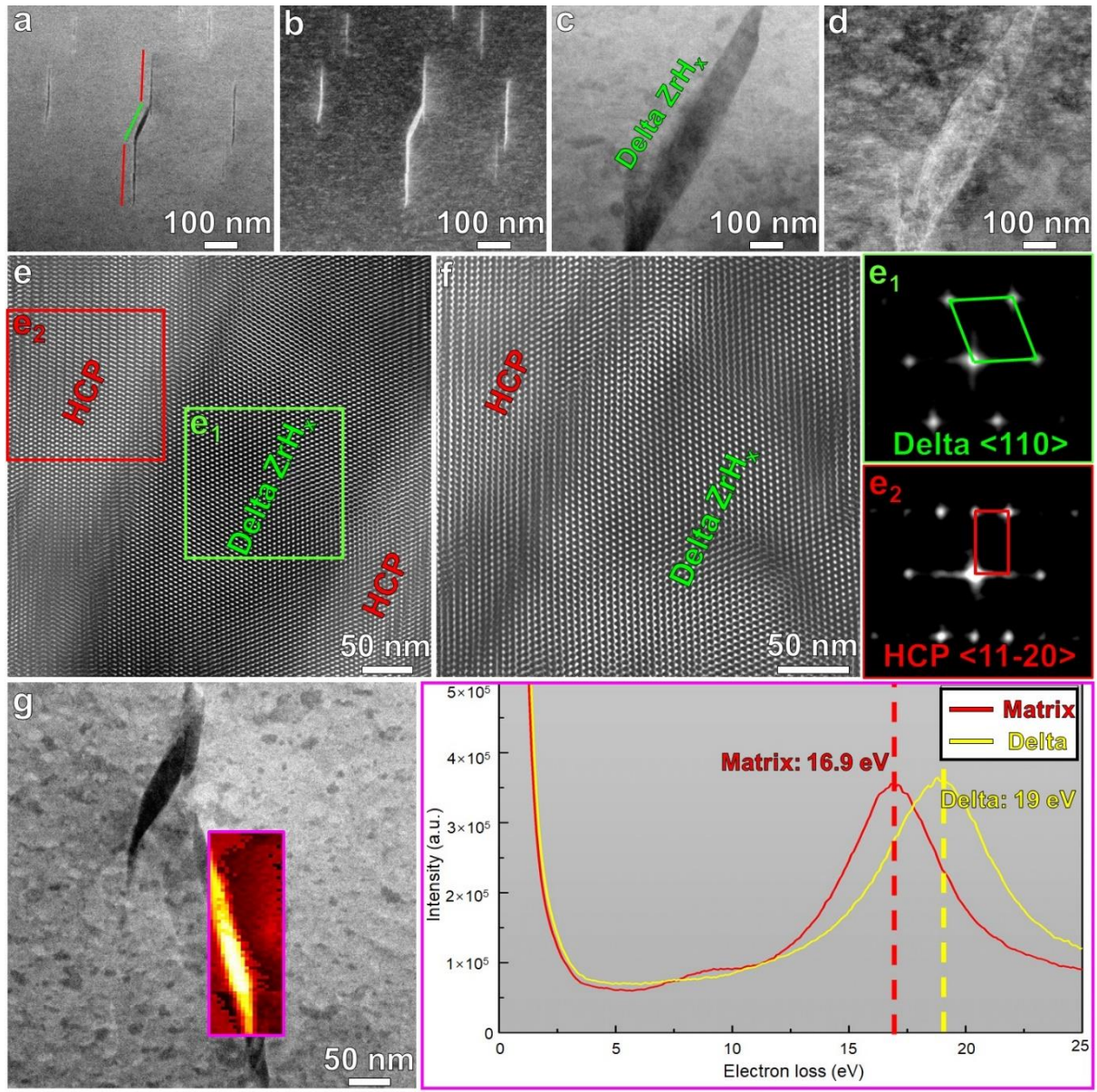

*Figure 4: The plate-shaped hydride along $\{1\bar{1}07\}$ habit plane (Type 2 highlighted by green line) is characterized at (a-d) low and high magnification by HAADF (a, c)- and LAADF (b,d)-STEM. Two Type 1 hydrides are connected by the Type 2 hydride. (e) and (f) High-resolution STEM images show the atomic column imaging of the domains, including pseudo diffraction patterns ($e_1$ and $e_2$) calculated from 2D-FFT of subregions within the atomic column images. (g) Area and corresponding EELS spectrum imaging overlaid phase map (yellow: hydride, red: matrix) confirming the Type 2 hydride is δ hydride.*

### APT analysis of a Type 1 Hydride and the Zr-matrix

Figure 5 presents results from APT characterisation of one of these small (we presume Type 1) hydrides contained within the Zr-matrix. The 3D reconstruction reveals a high aspect ratio region of high hydrogen concentration, consistent with similar observations in the STEM images. The concentration of hydrogen within the matrix is higher than an expected background [25], although due to instrument contributions the exact hydrogen concentration due to hydrogen stored in the Zr alpha matrix after quenching is unknown, but certainly greater than 0% and no more than 6 at%. The concentration of the hydrogen increases substantively within the hydride and reaches a maximum of ~17 at%. This is about half of the typically expected composition for a solid phase hydride, e.g. the lowest concentration of hydrogen in a hydride is found in the ζ-hydride phase of 33 at%, and this suggests that this is likely the precursor state of a hydride. Immediately on the side of the hydride domain, the Sn concentration is found to be enriched, hinting at rejection of Sn from the hydride region. The concentration of O is not correlated within the hydride. This Sn rejection is consistent with previous reports [42,43] , strengthening the hypothesis that this particle is indeed a hydride.

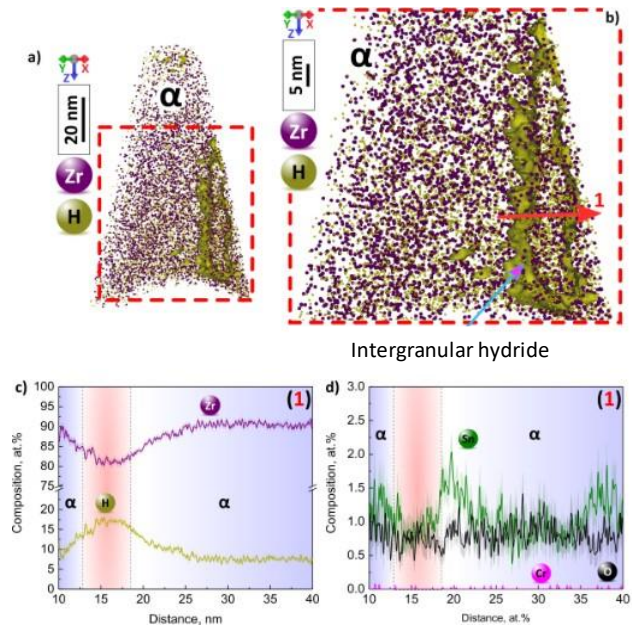

Intergranular hydride

*Figure 5:Atom probe tomography of a needle containing zirconium and hydrogen, showing (a) the reconstruction of the 3D needle, with (b) zoomed insert revealing a region of higher hydrogen concentration and the location of the concentration profile line used in (c) and (d). The concentration profiles of (c) Zr and H and (d) Sn and O for this region. Error bars are shown as lines filled with colour and correspond to the 2σ counting error.*

## Secondary Phase Particle

We explore the influence of an intermetallic SPP on the population of hydrides, as shown in Figure 6. From FFT analysis the particle is revealed to be FCC FeCrZr particle with an orientation relationship of $<110>_{\text{FCC}} || <11\bar{2}0>_{\text{HCP}}$ and the $\{2\bar{1}1\}_{\text{FCC}} || \{1\bar{1}01\}_{\text{HCP}}$ with the matrix.

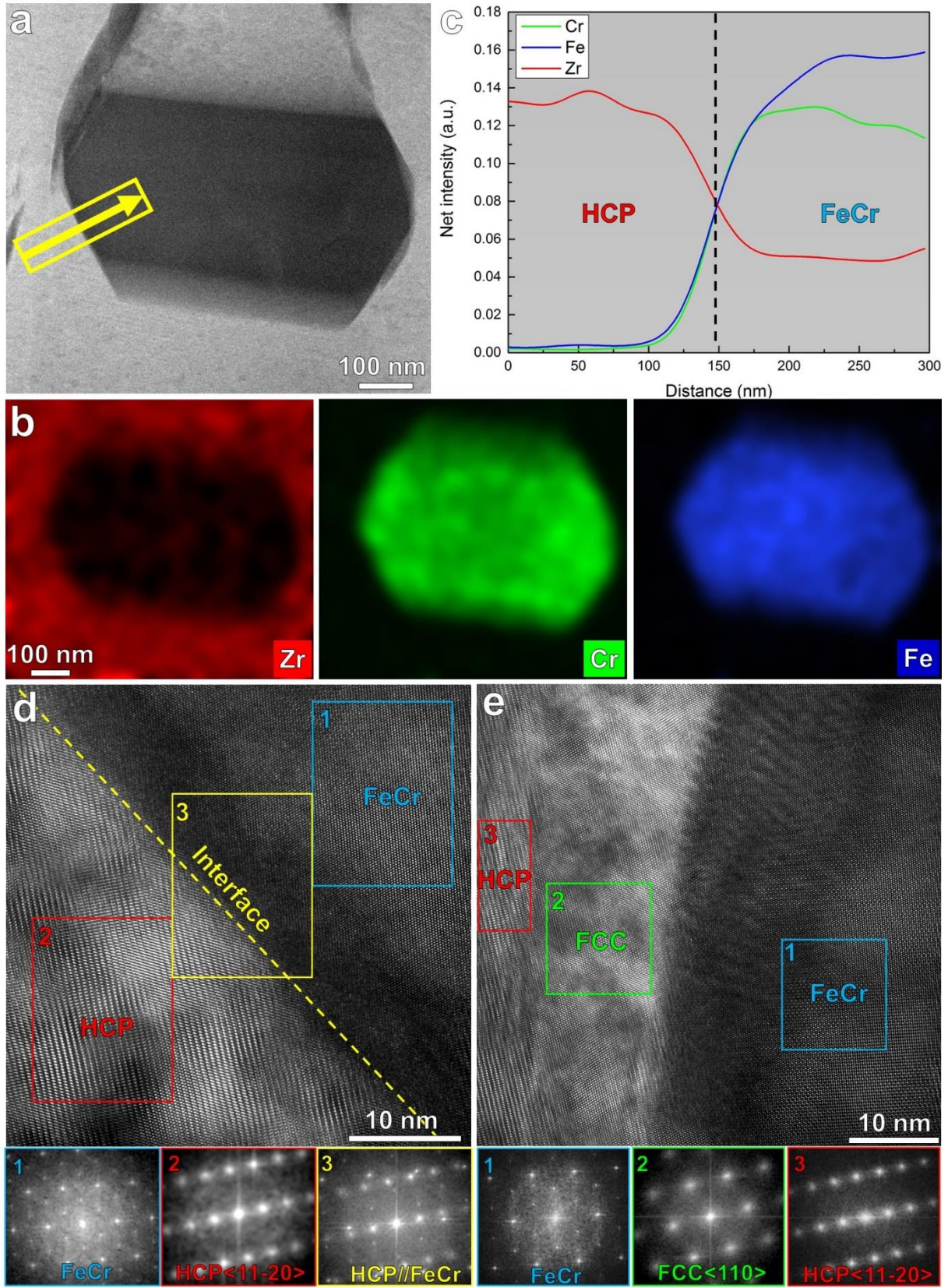

Figure 6: (a) HAADF-STEM imaging of a secondary phase particle within the matrix and (b) the corresponding EDS analysis of Zr, Cr and Fe. (c) 1D compositional profile within the yellow square highlighted in (a) along the direction marked with the yellow arrow. High resolution TEM image of the interface (d) between matrix and SPP, and (e) among matrix, hydride and SPP. The corresponding FFT patterns are shown below reveal the orientation relationship.

The SPP also changes how hydrides are formed within the matrix. Small hydrides are observed on the SPP-matrix interface. These hydrides have the same orientation relationship between the hydride and α-Zr matrix as all the other hydrides present in the TEM foil. The hydride decorates the interface of the SPP and Zr matrix, and the hydride looks to be only located within the Zr-matrix region.

## Discussion

This paper focusses on understanding the types and morphology of nanoscale hydrides, and the chemical segregation of hydrogen in the zirconium matrix. This is of interest when establishing the safety case for cladding, though we note that no irradiated samples have been explored in the present work.

Hydrogen is present within the matrix, i.e. not all the hydrogen migrates to form hydrides. This is shown within the atom probe analysis (Figure 5) that reveals up to 6 at.% H is present in the matrix phase. Even if this may be overestimated, and if the precise measurement of the H concentration by APT remains challenging for both Zr-hydrides[42] and the remaining content in the metallic matrix [44], the H-content in the matrix cannot be nill. We note that these samples were quenched in water after being held at 400°C and this could mimic conditions where a water quench is used to reduce the temperature of fuel rods. Local analysis in the matrix using APT presents an upper bound of 6 at.% H (1300 wppm) of hydrogen retained in the matrix and this compares well with prior density functional theory models by Lumley et al. [21]. Meanwhile, the hydride that is presented in Figure 5 has only 17 at% hydrogen compared to the minimum of 33 at% for the ζ-hydride phase. We believe this is the first time a hydride of stoichiometry less than ζ-phase has been observed. This suggests we have characterised a hydride in its embryonic early formation stage. The density functional theory based model by Lumley et al. suggest that below about 20 at% hydrogen, the hydrogen atoms want to repel and remain spaced out in the Zr lattice forming an extra barrier to hydride nucleation suggesting the hydride observed will be metastable.

In these samples, nanoscale hydrides are present with two forms. Type 1 hydrides form with their long axis indicating they are located on the basal plane. EELS analysis of the plasmon peak confirms they are γ-phase and the diffraction patterns of this phase are consistent with this analysis. These are different to the Type 2 hydrides which are inclined with respect to the Type 1; and tend to have a $\{1\bar{1}0\bar{7}\}_{\text{HCP}}$ interfacial plane, but they maintain a similar orientation relationship with the zirconium matrix. The analysed Type 2 hydride was confirmed to be δ-phase, using both EELS plasmon mapping and structural analysis.

Type 1 hydrides are commonly observed and tend to line up with each other in the matrix. They are very small (less than 10nm wide) and are unlikely to be observable with optical methods. Tin rejection ahead of the hydride-metal interface, into the matrix, is consistent with prior work [43] and confirms that these hydrides are formed during the quenching process, and not an artefact of the sample preparation route [40,45]. The interface of these hydrides is not flat and the termination of these hydride plates is not sharp (see Figure 3). They are often broken up and this may be important to understand hydride nucleation and growth.

Type 2 hydrides are observed linking Type 1 hydrides together. The Type 1 hydrides seem to extend beyond the edge of the Type 2 hydrides, as evidenced by the sharp tips of the upper Type 1 hydride in Figure 4b. The direction of the Type 2 hydride is likely directed by the combined misfit strain, and associated potential, of the two Type 1 hydrides in near proximity. If this is the case, there may be a combination of hydrogen concentration, Type 1 hydride distribution, and kinetics which will change the distribution of Type 1 and Type 2 hydrides at the nanoscale.

These results may be different to the prior work of Jia et al. [23], presumably due to the fact that the hydrides formed in the present study were created via redistribution following a dissolution above the hydrogen solvus line, and subsequent precipitation in Zircaloy-4. Our nano-scale analysis with combined TEM-based EELS and TEM-based diffraction confirms that we observe both fine scale Type 1 hydrides, as well as connecting Type 2 hydrides. Furthermore, both of these hydrides are typically much smaller than the hydrides characterised by Jia et al.

SPPs are intentionally used to improve the corrosion resistance of Zircaloy-4. This study only probes one SPP, and so we suggest care with extrapolation of our findings. In this case, we see that the SPP has a sharp interface with the matrix. The precise chemical composition of this SPP was not obtained, due to uncertainty in the EDS analysis and uncertainty on the through thickness ratio of α-Zr matrix to SPP, but we see that near atomically sharp interface with the matrix and that the SPP has facets with a six-fold symmetry in this projection. On one of these facets we can observe a hydride located between the SPP and the α-Zr matrix, and there is an orientation relationship between the hydride, SPP and matrix. The hydrides attached to the SPP appear to be consistent with Type 2 hydrides, and they bulge outwards (see Figure 6a) from the SPP indicating that the hydride grows first on the SPP-matrix interface, and further hydrogen thickens the hydride. This is consistent with Burr et al. [30] and Jones et al. [31] that H segregates to the SPP interface and perhaps can concentrate enough in that region to nucleate a hydride. The role of the SPP here is important to consider, as it may change the hydride distributions under irradiation as SPPs tend to dissolve under irradiation, with C-14 precipitates dissolving when held in service in a nuclear reactor for 10 years at 260-300°C, accumulating a neutron fluence of $10^{22}$ n/cm$^2$ [46].

# Conclusion

We characterise nanoscale hydrides in Zircaloy-4, formed *via* precipitation upon water quench after a thermal homogenisation of the hydrogen distribution. We observe two hydride types – a distribution of nm-sized Type 1 γ-phase hydrides with (0001) interfaces; a distribution of larger (but still ~nm-sized) Type 2 δ-phase hydrides with $\{10\bar{1}7\}$ interfaces bridging between Type 1 hydrides. The diffraction patterns of these two phases are very similar and could be confused, so we confirmed this with EELS analysis of the plasmon peak.

We also observe a Type 2 hydride formed on some of the facets of a FeCrZr SPP.

We observe Sn rejection ahead of the Type 1 hydride interface.

We also observe the presence of up to 6 at% hydrogen in the matrix local to the hydride interface, via APT, even when hydrides have been precipitated. This is in stark contrast to the 1 wppm solubility limit determined by global H measurements at room temperature.

The size, shape and distribution of these hydrides, and the presence of locally high hydrogen levels in the matrix, may influence the generation of physical models used to describe mechanical performance, as well as influence further hydrogen pick-up and corrosion of the fuel cladding material.

# Author Contributions

Wenjun Lu performed the transmission electron microscopy and associated analysis. Paris Kontis performed the atom probe tomography and associated analysis and scanning electron microscopy. Siyang Wang and Ruth Birch fabricated the samples and helped motivate the work. Mark Wenman helped motivate the work and provided valuable discussion. Baptiste Gault supervised the work and helped with the APT data processing and analysis. T. Ben Britton supervised the work and wrote the first draft of the manuscript. All authors contributed to the final manuscript.

## Acknowledgements

TBB, RB and SW acknowledge funding from Rolls-Royce plc and EPSRC for the MIDAS programme (EP/S01702X/1). TBB acknowledges funding of his research fellowship from the Royal Academy of Engineering. We thank Luca Reali for helpful discussions. We acknowledge Uwe Tezins, Chris Broß, and Andreas Sturm for their support to the FIB and APT facilities at MPIE. PK and BG are grateful for the Max-Planck Society and the BMBF for the funding of the Laplace and the UGSLIT projects respectively, for both instrumentation and personnel. BG is grateful for financial support from the ERC-CoG-SHINE-771602.

## Data Statement

<At proof stage – a completed zenodo bundle will be linked here>